# Quantum sensing with spin defects in boron nitride nanotubes

Roberto Rizzato[1,2]*, Andrea Alberdi Hidalgo[1], Linyan Nie[1,2], Elena Blundo[2,3], Nick R. von Grafenstein[1,2], Jonathan J. Finley[2,3], Dominik B. Bucher[1,2]*

[1] Technical University of Munich, TUM School of Natural Sciences, Chemistry Department, Lichtenbergstraße 4, Garching bei München, 85748, Germany
[2] Munich Center for Quantum Science and Technology (MCQST), Schellingstr. 4, München, 80799, Germany
[3] Walter Schottky Institute, TUM School of Natural Sciences, Physics Department, Am Coulombwall 4, Garching bei München, 85748, Germany

*Corresponding authors: roberto.rizzato@tum.de, dominik.bucher@tum.de

## Abstract

Spin defects in semiconductors are widely investigated for various applications in quantum sensing. Conventional host materials such as diamond and hexagonal boron nitride (hBN) provide bulk or low-dimensional platforms for optically addressable spin systems, but often lack the structural properties needed for chemical sensing. Here, we introduce a new class of quantum sensors based on naturally occurring spin defects in boron nitride nanotubes (BNNTs), which combine high surface area with omnidirectional spin control—key features for enhanced sensing performance. First, we present strong evidence that these defects are carbon-related, akin to recently identified centers in hBN, and demonstrate coherent spin control over ensembles embedded within dense, microscale BNNTs networks. Using dynamical decoupling, we enhance spin coherence times by a factor exceeding 300× and implement high-resolution detection of radiofrequency signals. By integrating the BNNT mesh sensor into a microfluidic platform we demonstrate chemical sensing of paramagnetic ions in solution, with detectable concentrations reaching levels nearly 1000 times lower than previously demonstrated using comparable hBN-based systems. This highly porous and flexible architecture positions BNNTs as a powerful new host material for quantum sensing.

## Main

Optically active spin defects in solid-state systems have emerged as fundamental tools in quantum technology and sensing[1–3]. The nitrogen-vacancy (NV) center in diamond[4,5] has become a well-established platform, enabling proof-of-concept breakthroughs including single-spin and single-molecule detection[6–14], high-resolution nuclear magnetic resonance (NMR)[15–18], and nanoscale sensing of chemical interactions at the liquid-solid interface[19–24] (Figure 1a). Additionally, extensive research has focused on leveraging the high, optically pumped electronic spin polarization of these systems to hyperpolarize nuclear spins[25–30], thereby enhancing signal strength in NMR spectroscopy[16,31]. Despite these advancements, it has become evident that the success of both sensing and hyperpolarization critically depends on maximizing the sensor–target interaction—a task that remains challenging for state-of-the-art solid-state systems[23,32,33]. Hexagonal boron nitride (hBN), a well-known two-dimensional (2D) van der Waals material (Figure 1b), has recently emerged as a promising alternative[34–39] due to the possibility of incorporating stable spin defects[34,40–44] into ultra-thin hBN flakes[45]. This enables the sensors to be positioned in proximity to the sample— with the potential to embed these nanoscale 2D sensors directly into target systems, e.g., layered heterostructures[40,44,46–54]. While this approach holds great promise for solid-state and materials research, it remains less suited for soft-matter environments and chemical sensing, where detecting molecular processes requires extremely intimate interaction and often multiple contacts between spin defects and analytes[55–59].

In this work, we introduce spin defects in boron nitride nanotubes (BNNTs)[60] (Figure 1c) as a quantum sensing platform for chemical analysis. In particular, we exploit two key features:

1) The intrinsic nanostructure of BNNTs enables mesh-like architectures—comprising nanotubes with hollow interiors and accessible surfaces[61–64], offering key advantages for applications requiring molecular confinement or high spin defect-to-target ratios[26,55–57,65–67].

2) Unlike NV centers in diamond or $V_B^-$ centers in hBN, which require a well-aligned bias magnetic field for qubit selection due to their preferential quantization axes[26,40,66–69], the spin defects in our BNNTs are optically active spin-½ systems which possess a unique isotropic magnetic response, allowing for omnidirectional spin control[41–43,60,70–73]. This is a key property for the development of high-surface-area sensors composed of randomly oriented nanoparticles (nanotubes), as it ensures that all embedded spin defects remain equally active regardless of their orientation.

We first characterize the properties of the spin ensembles and provide strong indications that they may correspond to the carbon-related $C_?$-defects recently identified in hBN crystals[70,72,73]. Then, we employ dynamical decoupling protocols to extend their room-temperature coherence times and implement state-of-the-art quantum sensing methods to achieve high-resolution radiowave frequency (RF) signal detection. Finally, we demonstrate the practical potential of this platform by integrating the BNNT-based mesh sensor into microfluidics. By monitoring the effect of paramagnetic ions on the spin properties of the $C_?$-defects, we successfully detect $Gd^{3+}$ ions at micromolar concentrations— approximately three orders of magnitude more sensitive than what has been reported using spin defects in hBN nanosheets[74,75]. These results highlight the versatility and promise of BNNT-based quantum sensing for applications in future chemical detection and hyperpolarization techniques.

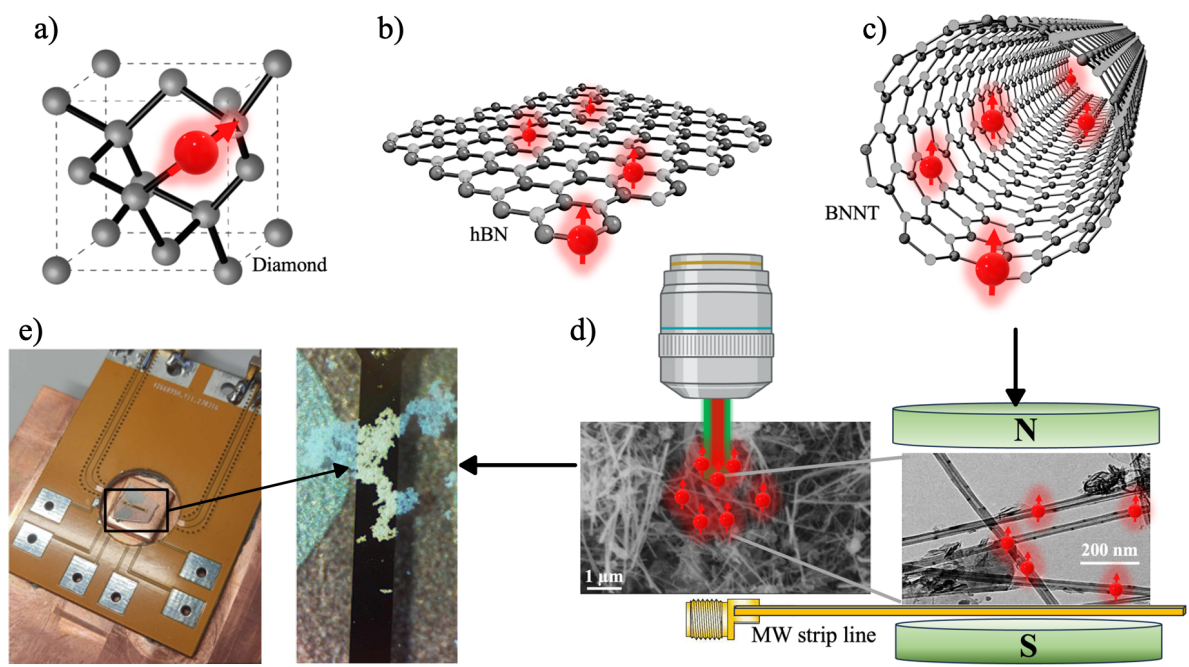

**Figure 1. Solid-state spins in semiconductor host materials.** Solid-state spin defects across different host materials, from **a)** bulk systems, e.g., NV centers in diamond, to **b)** two-dimensional (2D) materials, such as hexagonal boron nitride (hBN), and finally **c)** one-dimensional (1D) boron nitride nanotubes (BNNTs). **d)** BNNTs form a "mesh" architecture, characterized by a porous, high-surface-area structure suitable for chemical sensing and hyperpolarization applications. Electron microscopy images of our sample were kindly provided by the commercial supplier, NanoIntegris Inc. For optically detected magnetic resonance characterization, the BNNT mesh is positioned on a microwave strip line, allowing for efficient microwave (MW) delivery and optical excitation via laser illumination. **e)** Photographs of the MW delivery setup, with a close-up view of the area where BNNTs are drop-cast onto the gold strip line.

**Characterization of spin-defects in BNNTs**

We first characterized the spin defects naturally present in BNNTs using optically detected magnetic resonance (ODMR). All experiments were performed on BNNTs directly drop-cast on a microwave (MW) strip line (Figure 1e) under ambient conditions. Figure 2a presents ODMR spectra recorded at varying magnetic fields, revealing a single resonance line that shifts with the applied magnetic field and reaches a maximum contrast of 0.3–0.4% after optimizing MW and laser powers and pulse durations. Notably, the resonance remained identical and featureless regardless of the magnetic field orientation. Figure 2b shows the ODMR spectrum acquired using minimal laser and MW powers to reduce spectral broadening. The resonance is well fitted by a single Lorentzian with a full width at half maximum of ~ 25 MHz. This linewidth is consistent with that reported for single-spin defects in BNNTs[60] and about twice as narrow as that predicted for spin-pair complexes in hBN[73]. No significant inhomogeneous broadening—typically observed in ensembles—was detected, nor was hyperfine structure, suggesting that couplings to nearby nuclear spins (e.g., $^{11}B$ or $^{14}N$) are either weak or masked by the ~ 25 MHz linewidth.

We continue to demonstrate that we can efficiently control ensembles of spin defects in BNNTs under ambient conditions. This is achieved through Rabi oscillation shown in Figure 2c, recorded by driving the spins at the resonance frequency and varying the MW pulse duration from 0 to 200 ns. These measurements were performed using different MW powers, revealing the expected scaling of Rabi frequencies with the square root of the MW power (Fige 2d). Notably, at increased MW power levels, a distinct beating pattern emerges in the Fourier transform of the Rabi trace revealing two distinct peaks: a dominant component and a smaller one at twice the frequency (Figure 2e). A similar effect was recently observed by Scholten *et al.* for $C_?$-defects in hBN crystals, where it was attributed to weakly coupled spin-½ pairs[70]. The presence of this beating pattern suggests that we observe the same defect type, potentially consisting of electronic spin pairs that are collectively read out, as proposed in the hBN case. Furthermore, the positive contrast in the ODMR and its magnitude align well with *Optical-Spin Defect Pair* (OSDP) model recently proposed by Robertson *et al.*[73] for ensembles of $C_?$-defects in hBN crystals. According to the model, these spin defects may arise from electron hopping between two nearby defect sites, where optical activity occurs when both electrons are localized in a singlet state at the same site. Upon hopping, the electrons can form a weakly coupled radical pair, giving rise to both spin-parallel and spin-antiparallel states. When subjected to a bias magnetic field, these states are split by the Zeeman interaction and can be addressed by MW radiation.

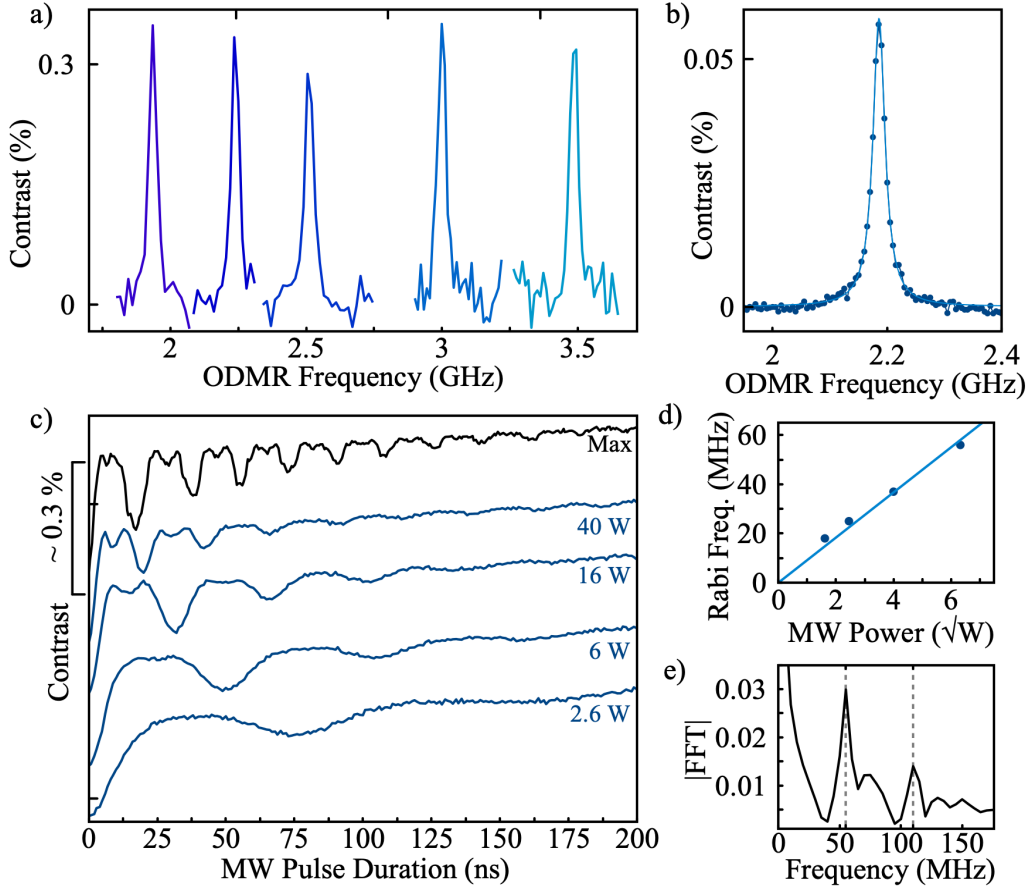

**Figure 2. Characterization of the spin defects in an ensemble of BNNTs. a)** ODMR spectra recorded at different bias magnetic field strengths. **b)** ODMR spectrum of BNNT spin defect ensembles at 78 mT. Blue dots show the experimental data, with a fitted Lorentzian line shape overlaid in light blue. **c)** Rabi oscillations recorded at various input MW powers (in Watt). The black line shows the experimental Rabi oscillation under optimized conditions of maximum input MW power showing a pronounced coherent beating pattern. **d)** Rabi frequency plotted as a function of the square root of the MW power, confirming the expected linear dependence. **e)** Fast Fourier transforms (FFT) of the Rabi oscillation at max MW power revealing two main frequency components: a dominant peak near 50 MHz and a smaller peak at twice that frequency, typically observed in spin pairs.

**Spin relaxation and coherence extension**

After obtaining strong indications of the presence of $C_?$-defects in BNNTs and demonstrating the coherent manipulation of the randomly oriented ensemble, we continue to explore advanced spin control. Firstly, we start with the measurement of the $T_1$ spin-lattice relaxation time (Figure 3a). We monitor the fluorescence contrast as the time interval T between two laser readouts increases. By fitting the decay, we extract a $T_1$ relaxation constant of approximately 26 µs. This value appears to be a factor of 2–3 longer than what has been reported for $C_?$-defects as single emitters in individual nanotubes and for defects in hBN crystals[60,70]. We then measured the native coherence time $T_2$ using a spin-echo sequence (Figure 3b). In this experiment, after initializing the spins with a laser pulse, a microwave π/2-pulse resonant with the $C_?$-defects' ODMR line creates spin coherence, which is refocused by a single π-pulse before being transformed back into spin populations by a final π/2-pulse. By monitoring the detected signal while varying the inter-pulse delay and, consequently, the total sequence time T, we recorded an exponential decay. Fitting this decay yields a $T_2$ time constant of approximately 50 ns, which is about half the value reported for single defects in individual BNNTs[60].

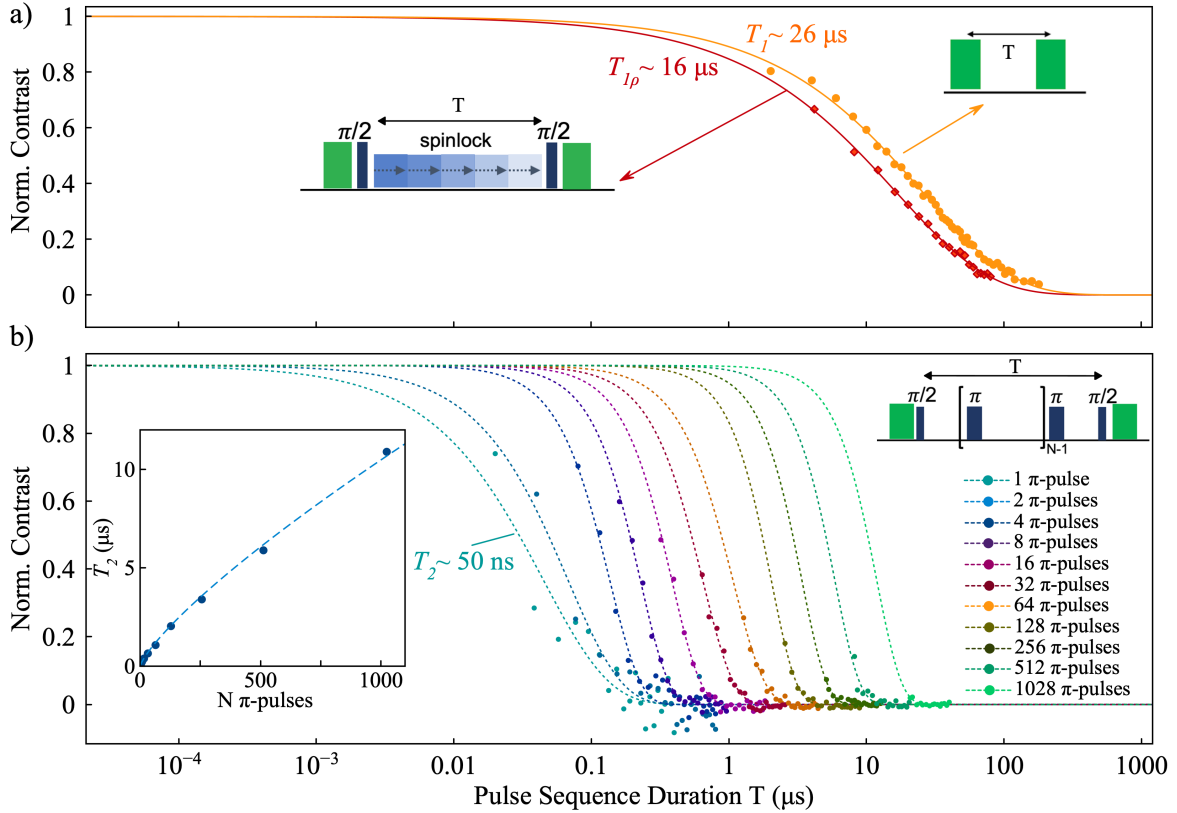

**Figure 3. Spin Relaxation and coherence extension of an ensemble of spin defects in BNNTs.** Semi-log plot of spin coherence and relaxation decays measured using various pulse sequences, illustrated in the insets. Colored dots represent experimental data, while lines correspond to fits. **a)** The solid orange data points and fit depict the fluorescence contrast decay for the $T_1$ spin-lattice relaxation measurement (sequence shown in the top-right inset). The solid red squares and line correspond to the coherence decay measured via spin-locking experiments, yielding $T_{1\rho}$ (pulse sequence shown in the inset on the left). **b)** The cyan dashed curve and dots show coherence decay measured using a simple Hahn-echo sequence (one π-pulse), from which the $T_2$ constant is extracted. All other dashed curves and colored dots represent coherence decays measured using CPMG sequences with an increasing number of π-pulses, $N$ (pulse sequence shown top-right inset). The bottom-left inset plots the coherence times $T_{2,\text{CPMG}}^{(N)}$ as a function of $N$, fitted to a power law model.

Furthermore, we present data on the extension of the coherence times via dynamical decoupling methods. In Figure 3b, data from multiple Carr-Purcell-Meiboom-Gill (CPMG) experiments are shown (colored dots and dotted lines). CPMG is a spin-echo based pulse scheme[1,53,76,77], that uses train of π-pulses for decoupling the spins from the environmental magnetic noise. This process effectively prolongs the $T_2$ of the BNNT spin defects and we can determine the corresponding increase in $T_{2,\text{CPMG}}^{(N)}$, with the number N of π-pulses. Remarkably, residual coherence is still observed even after ~$10^3$ π-pulses, where we reach coherence times of over 10 μs. We plot the coherence time as a function of the number of π-pulses $N$ in the inset in Figure 3b. It follows a sub-linear trend, which we fit with a power-law $T_2 = a\,N^s$, yielding an exponent $s \approx 0.79$. The extracted exponent is slightly higher than the theoretical scaling $T_2 \propto N^{2/3}$ expected for a Lorentzian spin bath in the slow-noise limit ($\tau_c \gg T$). Alongside the featureless ODMR spectrum – indicating weak coupling with nearby spins – this suggests deviations from an ideal Lorentzian model—possibly due to reduced bath spin density and more correlated (non-Markovian) dynamics[53,78–80].

Finally, we demonstrate coherence extension using the spinlock sequence[25,53,81]. In this method, after generating a superposition with a π/2 pulse, the spin is locked onto the axis by applying a longer MW pulse with a phase aligned with the spin vector. By monitoring the fluorescence contrast while sweeping the spin-lock pulse duration, we can measure the relaxation time in the rotating frame ($T_{1\rho}$) which is significantly longer than the native $T_2$ time, extending from ~ 50 ns up to ~ 16 μs and approaching the $T_1$ limit. This remarkable over-300-fold increase in coherence time highlights the effectiveness of the dynamical decoupling protocols in mitigating decoherence mechanisms of our randomly oriented BNNT ensemble. We note that the intricate Rabi dynamics of weakly coupled spin-½ pair ensembles posed challenges for precise microwave pulse calibration, highlighting the need for more systematic studies in future work. A description of the optimization procedure used in this study is provided in the Methods section.

**Sensing of RF signals**

Having demonstrated the effectiveness of dynamical decoupling protocols in enhancing coherence, we use our BNNTs spin defect ensemble to detect RF signals generated from an antenna placed nearby (Figure 4a). This is made possible by leveraging spin coherence, which enables the encoding of a relative phase in response to an external magnetic field. In the presence of oscillating magnetic fields—such as RF signals—this phase can be coherently accumulated using spin-echo-based techniques. By precisely tuning the timing of the pulse sequence (i.e., the delays between π-pulses) to match the oscillation period of the RF field, constructive phase buildup is achieved. This *tuning* allows the sensor's spin coherence to constructively integrate the field's effect, enabling high-sensitivity RF detection[1,53]. We used the Coherently Averaged Synchronized Readout (CASR)[15,53,82] which phase-locks a series of concatenated spin-echo sequences to the target RF field (Figure 4b). Each pulse sequence is tuned to the RF frequency and designed so that its total duration exactly matches an integer multiple of the RF wave's period. When fully matched, each sequence experiences the same phase of the RF field in every cycle, resulting in an identical fluorescence readout. However, when slightly detuned, the protocol effectively converts the RF signal into a low-frequency optical beating over time that encodes spectral information. By continuously collecting data over an extended period CASR achieves arbitrarily high frequency resolution, as its sensitivity is limited not by the spin coherence time, but by the timing stability of the control electronics. To demonstrate the implementation of this method for our randomly oriented BNNTs, we applied a 15 MHz RF signal and introduced a deliberate 1 kHz detuning for a total detection time of 2 seconds. We present the Fourier transform of the detected time-dependent signal in Figures 4b and 4c, yielding a sharp peak with a full-width at half-maximum of ~1 Hz, as shown in Figure 4d. Additionally, we employed the same CASR protocol to evaluate the achievable sensitivity of the mesh sensors, obtaining a sensitivity of approximately 20 μT/√(Hz). This sensitivity is approximately ten times lower than that demonstrated using ensembles of $V_B^-$ centers in hBN[53] and does not yet match the performance of established quantum sensing platforms. However, it is important to emphasize that this system is still in its infancy. The current results were achieved using naturally occurring spin defects in as-received BNNT material, without any form of defect engineering or material optimization. Importantly, the ability to apply advanced quantum control protocols—such as NOVEL[27,83] or PulsePol[30,66]—to spin defects in a randomly oriented host paves the way for implementing solid-state, spin-based hyperpolarization techniques.

**Sensing of paramagnetic ions**

Spin-defect-based quantum sensors offer a powerful platform for detecting paramagnetic ions at the nanoscale, with key applications in chemical, biological, and medical sciences[23,69,75,84–88]. This functionality is enabled by the sensitivity of the spin defect's $T_1$ to magnetic noise generated by paramagnetic species in the nearby environment. Building on this principle, we demonstrate the first practical realization that fully embodies the original vision behind our BNNT mesh sensor—combining high surface area[89] and omnidirectional spin response with straightforward integration into a microfluidic sample-handling system[90]. To achieve this, we designed the configuration shown in Figure 4e, where a microfluidic channel with an open bottom was placed on top the MW strip line, where the BNNTs were drop cast (see Methods), ensuring direct contact between the liquid sample and the mesh sensor. A solution with varying concentrations of $Gd^{3+}$ ions was introduced, and their presence was successfully detected by monitoring changes in the spin properties of the BNNT based spin defects (Figure 4f). When in contact with pure distilled water, the defects exhibit a $T_1$ time of approximately 27 μs, consistent with the value observed in our previous experiments on BNNT samples in air. Upon exposure to a 1 mM solution of $Gd^{3+}$ ions, the $T_1$ time significantly decreases, along with a reduction in ODMR contrast (Figure 4f, inset). The direct correlation between the two experiments arises because both the optical initialization/readout and the microwave pulses in our ODMR protocol last 5 μs each, resulting in a total sequence duration that is on the order of the $T_1$ timescale. When $T_1$ shortens, as in the presence of $Gd^{3+}$ ions, spin populations relax more rapidly during the measurement, reducing the spin-state contrast and thus the ODMR signal[75]. This makes ODMR contrast an effective and convenient proxy for detecting changes in relaxation dynamics. The BNNT mesh sensor can be regenerated by flushing the system with an acidic solution (pH ~ 2), followed by rinsing with pure water, which fully restores its performance. Finally, we examined the impact of paramagnetic $Gd^{3+}$ ions on the ODMR contrast (Figure 4g), revealing a cooperative-binding-like interaction where local magnetic noise from $Gd^{3+}$ ions induces saturable contrast quenching (fit in Figure 4g). The fit yields a dissociation constant of $K_d \approx 17$ μM, indicating that half of the maximal quenching occurs at this concentration. Furthermore, the nonzero baseline suggests that a subpopulation of spin defects remains unaffected, likely due to their buried location within the BNNT nanotube (see Methods for details). While a direct comparison with previously reported systems is challenging—due to variations in structural architectures, defect types, and experimental conditions—our results underscore the distinct advantages of the BNNT mesh sensor, which enables detection of paramagnetic ions in the low micromolar range. This detection capability exceeds by several orders of magnitude what has recently been demonstrated with $V_B^-$ defects in hBN-based platforms[74,75]. Remarkably, this enhancement is achieved despite those systems exhibiting superior intrinsic spin properties—such as up to ~100× higher ODMR contrast and comparable relaxation times. This indicates that the improved performance of our BNNT sensor is likely driven by its unique architecture and high surface accessibility, rather than the intrinsic qualities of the spin defects themselves. These findings highlight the promise of BNNT-based quantum sensors for chemical sensing applications.

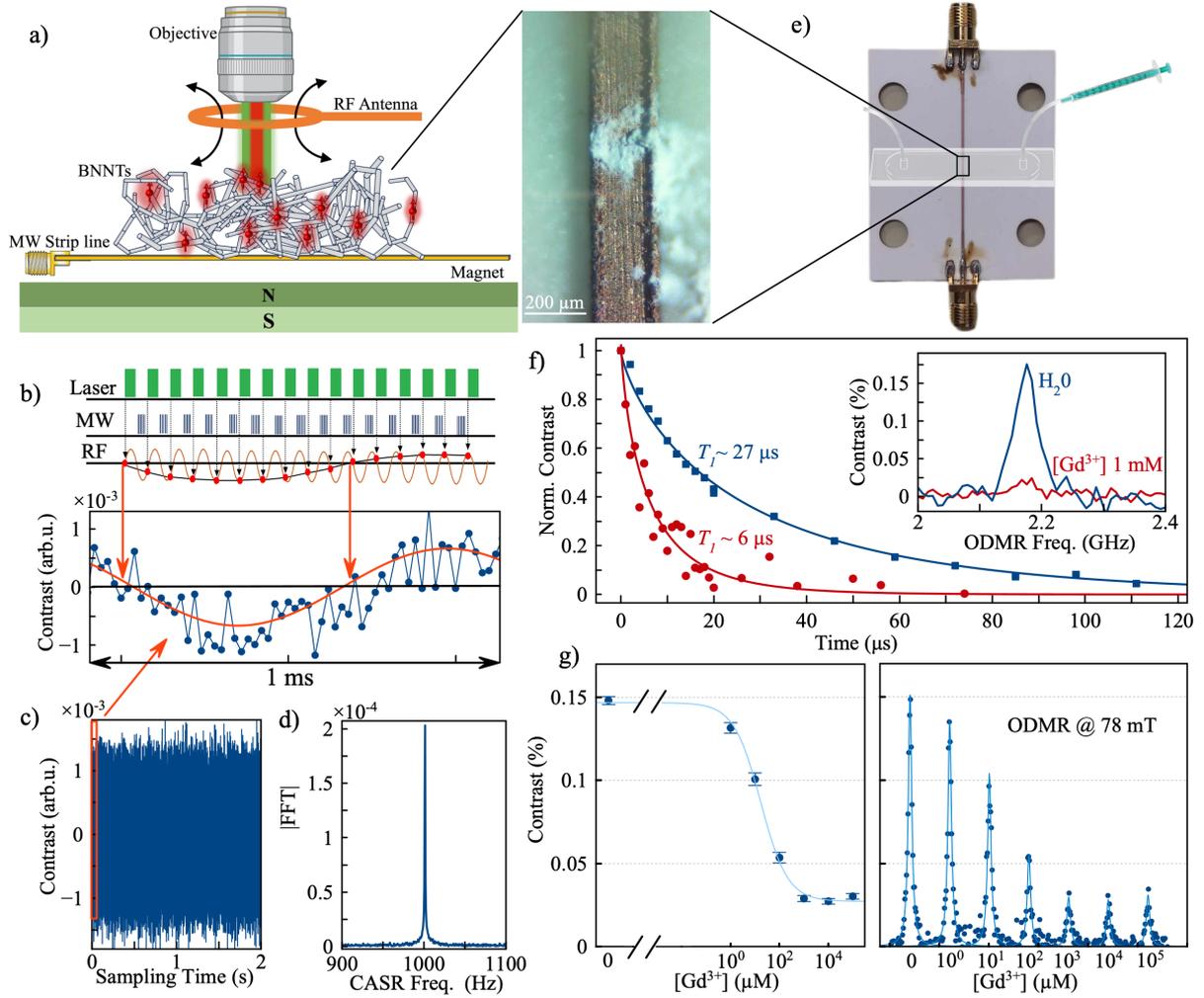

**Figure 4. Quantum sensing with ensembles of spin defects in BNNTs. a)** Simplified schematic of our setup for the detection of RF signals through the BNNTs mesh sensor. **b)** Top: Coherently Averaged Synchronized Readout (CASR) pulse sequence. 2-π pulses-spin echo subsequences are synchronized to the RF sensing frequency for 2 seconds. Each detected point corresponds to an optical readout, represented by the red circles. Bottom: Zoom-in of the experimental CASR time. **c)** CASR time trace. **d)** Fourier transformation of the time trace in c) revealing a sharp peak at the CASR Frequency. **e)** Integrated microfluidic chip for in-solution quantum sensing using a BNNTs mesh. A copper MW strip line enables coherent spin control, while the microfluidic chip—open at its bottom—is directly mounted on top of the strip line and sealed. Inlet and outlet tubing enables solution flow across the sensing region. The zoom-in highlights the area where BNNTs are deposited onto the strip line, appearing as a porous, textured white powder. **f)** $T_1$ relaxation curves of spin defects in BNNTs exposed to distilled water (blue) and to a 1 mM solution of $Gd^{3+}$ ions (red), along with corresponding exponential fits. The presence of paramagnetic ions results in a pronounced reduction in spin relaxation time. Inset: Corresponding ODMR spectra under both conditions. **g)** $Gd^{3+}$ concentration dependence of the ODMR contrast. Left: Semilog plot of the maximum ODMR contrast as a function of $Gd^{3+}$ concentration. The data are well fit by a Hill-like function (light blue), indicating a progressive and saturable decrease in contrast. Right: Representative ODMR spectra acquired at $B_0 = 78$ mT, corresponding to the data points shown on the left, illustrating the progressive reduction in signal intensity with increasing $Gd^{3+}$ concentration.

## Outlook

We present a fundamentally new platform—boron nitride nanotubes (BNNTs)—for spin defect-based quantum sensing. This system overcomes the constraints of traditional bulk and planar quantum materials by introducing a nanoporous, high-surface-area architecture that enhances analyte interactions while preserving robust room-temperature quantum control. At the heart of this system are optically active spin-½ defects, which we attribute to spin-pair $C_?$-defects, previously reported in hBN flakes[70,72,73]. We demonstrated coherent control of randomly oriented nanotube ensembles, achieved a two-order-of-magnitude increase in coherence time via dynamical decoupling protocols, and demonstrated their ability to detect oscillating magnetic fields in the RF frequency range. However, the platform's potential is best exemplified through its integration with microfluidics, where the BNNTs' nanoporous geometry enables efficient analyte interfacing allowing us to detect paramagnetic $Gd^{3+}$ ions down to low micromolar concentrations. This corresponds to a roughly 1000-fold lower detectable concentration compared to hBN-based sensors[74,75], highlighting the critical role of sensor architecture and surface accessibility in enabling effective detection of paramagnetic species. Combined with advanced spin control capabilities, this system offers a promising route toward efficient polarization transfer for spin-based optical hyperpolarization, enabling enhanced sensitivity in NMR spectroscopy[26,28,91,92]. Finally, BNNTs can act as nanoconfined reaction chambers with built-in quantum sensors, providing a powerful platform for monitoring chemical processes at the nanoscale and single-molecule studies[8].

## Methods

### Experimental Setup

Initialization of spin defect ensemble is realized with a 520 nm laser (Cobolt, 06-01) at a power of approximately 150 mW (continuous wave). The excitation laser light is reflected by a dichroic mirror (DMLP550, Thorlabs), after which it is focused on the BNNTs sample by an objective (CFI Plan Apochromat VC 20×, NIKON) with a numerical aperture (NA) of 0.75 and generating a laser spot size of around 20 μm diameter. For the paramagnetic ion sensing experiment, a larger working distance (5 mm) objective was used (100X Nikon CFI60 TU Plan Epi ELWD Infinity). Photoluminescence (PL) is collected by the same objective and alternatively focused by a tube lens on either: (1) an avalanche photodiode (APD) (A-Cube-S3000-03, Laser Components) for the spectroscopic path, or (2) a camera (a2A3840-45ucBAS, Basler) for imaging the sample. The excitation green light and unwanted fluorescence from other defects are filtered out using a long-pass filter with a cut-on wavelength of 550 nm (FEL0550,Thorlabs). The output voltage of the APD is digitized using a data acquisition unit (USB-6221 DAQ, National Instruments). An arbitrary waveform generator (AWG) with up to 2.5 GS/s (AT-AWG-GS2500, Active Technology) is used to synchronize the experiment and generate MW pulse sequences at 250 MHz pulses for spin control, which are mixed with a signal of a local oscillator generator (SG384, Stanford Research Systems) and an IQ mixer (MMIQ0218LXPC 2030, Marki) to reach the final frequency (~2 GHz). The microwave signal is pre-amplified (Mini-Circuits ZX60-153LN-S+) and then passed through a high-power amplifier (ZHL-100W-242+, Mini-Circuits). MW delivery to the BNNTs sample is achieved via a gold strip line, or via a copper strip line in the case of paramagnetic ion sensing. The copper strip line was specifically designed to interface easily with a microfluidic chip, featuring an uninterrupted substrate without vias and cutouts and a mildly textured conductor surface that promotes strong adhesion of the BNNTs, ensuring they remain anchored even under liquid flow. A permanent magnet beneath the sample holder provides a magnetic field in the range of ~50–200 mT. RF signals for AC magnetometry are generated by a waveform generator (DG1022Z, Rigol) connected to a power amplifier (N-DP340, PRANA). A small wire loop was placed near the sample to deliver RF signals.

**MW delivery setup for characterization and RF sensing.** MW strip lines (as shown in Figure 1e) were designed and fabricated in-house via optical lithography and e-beam Au evaporation to provide microwave fields for coherent electron spin manipulation. Either linear strip lines or omega-shaped strip lines were realized. The microstrip layouts were optimized to maintain impedance matching, with a characteristic impedance of ~ 50 Ω. The linear strip lines were characterized by a length of 4 mm, and width of 350 μm, while the omega-shaped strip lines were characterized by a ring with inner radius of 500 μm and outer radius of 650 μm (width 150 μm); the ring is interrupted by a 400 μm spacing connected to linear legs tangential to the ring with length of 2 mm. For the realization of the strip lines, optical lithography with a maskless aligner by Heidelberg Instruments was employed, and 10 nm of Ti + 200 nm of Au were evaporated with a Lesker e-beam evaporator on a one-side-polished 350-μm-thick sapphire substrate to enable thermal dissipation. The chip was glued by thermally conductive paint on a ~1-cm-thick copper block, chosen for its high thermal conductivity. A printed circuit board (PCB) with outer dimensions of 5 cm × 4 cm and an inner hole with diameter of about 1.4 cm was mounted on the copper block with the chip being centered on the hole, and the strip line was Au-bonded to the PCB.

**MW delivery setup for paramagnetic ions sensing.** A MW strip line (shown in Figure 4a) was designed in-house and fabricated (CERcuits, Geel, Belgium) to provide MW fields for coherent electron spin manipulation. The microstrip structure was implemented on a printed circuit board (PCB) with outer dimensions of 50.6 mm × 40.6 mm. The substrate material was alumina ($Al_2O_3$) with a thickness of 500 μm, chosen for its high thermal conductivity. The conductive layer and the ground plane were both composed of copper with a thickness of 70 μm. The signal conductor was realized as a straight line with a width of 480 μm, resulting in a characteristic impedance of ~ 50 Ω. To enhance the local MW field strength at the region of the sample, a ~ 2 mm-long constriction was introduced at the center of the line, reducing the width to 240 μm. This narrowed region serves to locally increase the current density and thereby intensify the MW magnetic field in the vicinity of the sample positioned above it. The microstrip layout was optimized to maintain impedance matching and minimize signal reflection while providing sufficient field strength for fast coherent spin-state manipulation.

**Sample preparation.** The boron nitride nanotubes (BNNTs) were purchased from Nanointegris Inc. Throughout this study, the spin defects utilized were naturally present within the sample, which was used without further modification. The BNNTs were transferred onto the devices for MW delivery by dispersing them in ethanol and drop casting them onto the strip line using a pipette, followed by solvent evaporation.

**Spin defects characterization**

**ODMR measurements.** The spin defects were optically excited using a 5 μs-long laser pulse, and the electron spin resonance (ESR) transition driven using a 5 μs MW pulse at ~ 5 mW power. Fluorescence contrast was monitored by comparing the signal sequence with a reference sequence (MW off), applied immediately afterward for normalization and noise suppression[93]. Each spectrum in Figure 2a was recorded at different magnetic field strengths by adjusting the distance of a permanent magnet. The measurement sequence consisted of a 100 μs laser pulse followed by a 10 μs MW pulse. Each spectrum was acquired recording 10,000 averages per frequency point. In the spectrum of Figure 2b, a total of 100,000 averages per frequency point and 26 averages of the full sweep were recorded. The spectrum was fitted with a Lorentzian function: $L(f) = \frac{A}{1+\left(\frac{f-f_0}{\Delta f}\right)^2}$, where $A$ is the amplitude, $f_0$ the resonance frequency, and $\Delta f$ the half-width at half-maximum (HWHM). The fitted parameters are: $A = 5.83 \times 10^{-4}$; $f_0$ = 2.19 GHz, and HWHM = 12.9 MHz.

**Rabi experiments.** Rabi oscillations were recorded by setting the MW frequency to the center of the ODMR spectrum (2.274 GHz) and incrementally sweeping the MW pulse duration. The amplitude of the local oscillator was varied between datasets to generate Rabi oscillations at different MW powers. Considering the gains from the preamplifier (18.7 dBm) and the final outpower of the amplifier (50 dBm), the total applied MW power across datasets ranged approximately from 2 W to 40 W, as shown in Figure 2c. Each data point represents an average over 410,000 repetitions, using the same normalization and noise cancellation protocol previously described. The experimental Rabi traces were Fourier transformed to (1) extract the precise Rabi frequencies and (2) reveal the presence of a characteristic beating pattern. To examine the dependence of the Rabi frequency on MW power, we extracted the Rabi frequencies from the Fourier transforms of the oscillations (using the main peak) and plotted them as a function of the square root of the MW power. A linear fit of the form $f(x) = ax$ yielded $a = 9.2$ MHz/$\sqrt{W}$ (Figure 2d).

**T₁ measurement.** The longitudinal relaxation time $T_1$ (see Figure 3a) was measured by varying the delay time T between the initialization and readout laser pulses over a range from 2 μs to ~180 μs. To improve signal stability and suppress noise, a reference sequence was applied immediately after each measurement sequence, differing only by the inclusion of a MW π-pulse after the first laser pulse. Each data point was obtained by dividing the fluorescence readouts of the measurement and reference sequences. Every point was averaged 10,000 times, and the full time-sweep was repeated and averaged over 50 iterations. The resulting decay curve was fitted using a stretched-exponential function of the form: $f(t) = exp\left[-(t/T_1)^c\right]$. The fit yielded $T_1 = 25.87 \pm 1.03$ μs and $c = 0.665 \pm 0.008$.

**T₁ρ measurement.** The rotating frame spin-lattice relaxation time $T_{1\rho}$ was measured using a pulse sequence [(π/2)$_x$–d–(spinlock)$_y$–d–(π/2)$_x$] and monitoring the resulting fluorescence contrast while applying step-by-step increments of the spinlock pulse duration. 5 ns-long π/2-pulses were used and the delay times $d$ was kept to a value as short as possible (~1–2 ns). The amplitude of the spin-lock pulse was set to approximately 30% of the MW amplitude used for the π/2 pulses, and the MW frequency was detuned by 40 MHz from the exact resonance, as these conditions gave the best signal contrast in our experiments. Each data point was averaged over 100,000 repetitions to improve signal-to-noise ratio. The resulting curve fits a stretched exponential decay of the form: $f(t) = a \cdot exp\left[-(t/T_{1\rho})^c\right]$. The fit yielded $T_{1\rho} = 16.57 \pm 1.66$ μs and a stretch factor $c = 0.640 \pm 0.036$, with $a = 1$ fixed.

**T₂ measurements.** The spin-echo sequence followed the standard protocol [(π/2)$_y$–τ–(π)$_x$–τ–(π/2)$_y$], where the interpulse delay τ was swept from 10 ns to ~ 200 ns. To enable noise cancellation, a referencing scheme was employed by alternating the final (π/2)$_y$ pulse with a (3π/2)$_y$ pulse in successive measurements. Each data point was averaged 10,000 times, and the full sweep over τ values was repeated 100 times for improved signal-to-noise. The resulting decay curve was fitted using a stretched exponential function of the form: $f(t) = a \cdot exp\left[-(t/T_2)^c\right]$, with $a = 1$ fixed. The fit yielded a coherence time $T_2 = 44.71 \pm 4.7$ ns and a stretch factor $c = 0.902 \pm 0.133$.

**CPMG measurements.** We employed the same general approach used for the $T_2$ measurement—monitoring the spin-echo signal while increasing the free evolution time τ. Multiple decoherence curves were acquired for pulse sequences containing an increasing number $N$ of π-pulses, up to $N$=1000. To enable direct comparison across datasets, a consistent normalization procedure was applied. First, the time axis of each dataset was rescaled by a factor $S = 2N$, corresponding to the total evolution time $T$ between the initial and final π/2 pulses in the respective sequences. Next, the decay curve from the standard spin-echo experiment (i.e., $N = 1$) was fitted with the function: $f(T) = A \cdot exp\left[-(T/T_2)^c\right]$. The amplitude $A$ yields the spin-echo contrast corresponding to $T = 0$. All datasets were normalized to this value, and fitted using the same function, with the amplitudes kept to $A = 1$ (Table 1). The dependence of the coherence time $T_2$ on the number of π-pulses, as shown in the inset of Fig. 2b, was fitted with a power-law function of the form $f(N) = a \cdot N^S$, yielding fit parameters $a = (33\pm2)$ ns and s=0.79±0.01.

Table 1. Fitted T₂ times and stretched exponents c for different numbers of π-pulses N in the dynamical decoupling sequence.

| N (π-pulses) | T₂ (ns) ± SD | Stretch exponent c ± SD |
|---|---|---|
| 1 | 45 ± 5 | 0.90 ± 0.13 |
| 2 | 68 ± 5 | 1.11 ± 0.15 |
| 4 | 136 ± 3 | 1.93 ± 0.15 |
| 8 | 227 ± 2 | 2.09 ± 0.08 |
| 16 | 384 ± 4 | 2.07 ± 0.09 |
| 32 | 644 ± 10 | 1.93 ± 0.10 |
| 64 | 1069 ± 23 | 1.87 ± 0.11 |
| 128 | 2032 ± 48 | 2.39 ± 0.17 |
| 256 | 3391 ± 35 | 2.40 (fixed) |
| 512 | 5885 ± 106 | 2.40 (fixed) |
| 1024 | 11871 ± 183 | 2.40 (fixed) |

**Calibration of the MW pulses**. Optimal conditions for CPMG and spin-lock experiments were not achieved by simply driving the spins precisely on resonance or by selecting π/2 and π pulse durations based on Rabi revivals or their Fourier transforms. Instead, we found that more reliable results were obtained by fixing the pulse durations (e.g., 5 ns and 10 ns for the π/2 and π pulses, respectively) and empirically tuning the local oscillator amplitude to maximize the spin echo signal. In the case of spin-lock experiments, we consistently observed enhanced signal contrast when the microwave drive was deliberately set slightly off-resonance.

**RF sensing.** We applied a pulse sequence synchronized with the sensing RF signal, consisting of concatenated 2-π-pulse spin echo (XY-2) subsequences. These subsequences were repeated such that the timing between them matched an integer multiple of the RF period. Each XY-2 sequence included 5 μs laser pulses for initialization and readout, and 5 and 10 ns as π/2- and π-pulses, respectively, for MW manipulation. The inter π-pulse delay was set to τ = 16 ns corresponding to a target RF frequency of $\nu_{RF}$ = 15.625 MHz. For generating a relative frequency offset Δν = 1000 Hz, the RF frequency was shifted slightly to $\nu_{RF}$ = 15.626 MHz. The total sampling time was $t_s$ = 2 seconds. The resulting time-domain signal (shown in Figures 4b and 4c) was Fourier transformed, and the absolute value was plotted in Figure 4d, revealing a narrow peak with a full width at half maximum of approximately 1 Hz. The final signal was obtained after averaging 1000 repetitions. Using the calibration procedure described in Ref.[53], we estimated a magnetic field sensitivity of ~ 20 μT/√Hz.

**Paramagnetic ion sensing.** A small drop of the BNNT dispersion—prepared as described in the Sample Preparation section—was drop-cast onto the microwave strip line and allowed to dry completely as the solvent evaporated. This strip line was then glued to a custom-made glass microcuvette, designed in-house to enable continuous liquid flow through the channel (Figure 4e). ODMR and $T_1$ relaxation measurements were performed using the protocols described above, with the BNNT mesh sensor in contact with either pure water or $Gd^{3+}$ ion solutions at increasing concentrations, ranging from 1μM to 100 mM. Between measurement sessions, the sensor was reused following a cleaning procedure that involved flushing with several milliliters of distilled water, an acidic wash using $H_2SO_4$ at pH 2, and a final rinse with pure distilled water. For the ODMR data shown in Figure 4f, $T_1$ measurements were obtained using 20,000 averages and 100 sweeps for the $Gd^{3+}$ solution, and 20,000 averages with 20 sweeps for pure water. The corresponding ODMR spectra in the inset were acquired by averaging 20,000 repetitions per point and performing three full spectral sweeps for pure water, and 100 sweeps for the 1 mM $Gd^{3+}$ solution. In the data presented in Figure 4g, the ODMR signal was monitored while $Gd^{3+}$ solutions of increasing concentration (from 1 μM to 100 mM) were injected into the microchannel. At each concentration step, the channel was first flushed with 1 milliliter of the respective solution to allow the sensor to equilibrate. Each ODMR spectrum was then recorded using 20,000 averages and 9 sweeps (~ 4 minutes of acquisition time per spectrum), except for the last three concentrations (from 1 mM to 100 mM), for which 48 sweeps (~ 20 minutes acquisition time) were used to improve signal quality. After fitting each spectrum with a Lorentzian function, the maximum ODMR contrast was plotted as a function of $Gd^{3+}$ concentration to extract the concentration dependence. The data were well described by a generalized Langmuir (or Hill-like) function of the form $C_0/(1 + ([Gd^{3+}]/K_d)^n) + A$, where $C_0$ is the unperturbed ODMR contrast (distilled water), $K_d$ the dissociation constant, $n$ is the Hill coefficient, and $A$ is a saturation offset accounting for the nonzero baseline contrast observed at high $Gd^{3+}$ concentrations. The best-fit parameters were: $C_0$ = 0.1195 ± 0.0045, $K_d$ = 1.69×10$^{-5}$ ± 4.20×10$^{-6}$, $n$ = 0.78 ± 0.11, $A$ = 0.0272 ± 0.0020.

**Author Contribution.** R.R. and D.B.B. conceived the idea and designed the research. R.R. carried out the experiments and simulations. R.R. and A.A.H. built the experimental setup, with L.N. contributing to its optimization. E.B. provided the microstructure for microwave delivery used in the characterization, coherence extension, and RF sensing studies. N.vG. designed and supplied the structure for microwave delivery used in the ion sensing experiments. R.R. and L.N. implemented the integration of the BNNTs sensor with microfluidics. J.J.F. provided guidance on theoretical and experimental aspects. R.R. and D.B.B. analyzed the data. R.R. and D.B.B wrote the manuscript with input from all authors.

**Acknowledgements.** This project has been funded by the Bayerisches Staatministerium für Wissenschaft und Kunst through project IQSense via the Munich Quantum Valley (MQV), the Deutsche Forschungsgemeinschaft (DFG, German Research Foundation)—412351169 within the Emmy Noether program and the European Research Council (ERC) under the European Union's Horizon 2020 research and innovation programme (Grant Agreement No. 948049). The authors acknowledge support and seed funding by the DFG under Germany's Excellence Strategy–EXC 2089/1-390776260 and the EXC-2111 390814868.

**References**

1. Degen, C. L., Reinhard, F. & Cappellaro, P. Quantum sensing. *Rev. Mod. Phys.* **89**, 035002 (2017).
2. Awschalom, D. D., Hanson, R., Wrachtrup, J. & Zhou, B. B. Quantum technologies with optically interfaced solid-state spins. *Nature Photon* **12**, 516–527 (2018).
3. Wolfowicz, G. *et al.* Quantum guidelines for solid-state spin defects. *Nat Rev Mater* **6**, 906–925 (2021).
4. Balasubramanian, G. *et al.* Nanoscale imaging magnetometry with diamond spins under ambient conditions. *Nature* **455**, 648–651 (2008).
5. Doherty, M. W., Manson, N. B., Delaney, P. & Hollenberg, L. C. The negatively charged nitrogen-vacancy centre in diamond: the electronic solution. *New Journal of Physics* **13**, 025019 (2011).
6. Lovchinsky, I. *et al.* Nuclear magnetic resonance detection and spectroscopy of single proteins using quantum logic. *Science* **351**, 836–841 (2016).
7. Shi, F. *et al.* Single-protein spin resonance spectroscopy under ambient conditions. *Science* **347**, 1135–1138 (2015).
8. Du, J., Shi, F., Kong, X., Jelezko, F. & Wrachtrup, J. Single-molecule scale magnetic resonance spectroscopy using quantum diamond sensors. *Rev. Mod. Phys.* **96**, 025001 (2024).
9. Staudacher, T. *et al.* Nuclear Magnetic Resonance Spectroscopy on a (5-Nanometer)$^3$ Sample Volume. *Science* **339**, 561–563 (2013).
10. Mamin, H. J. *et al.* Nanoscale Nuclear Magnetic Resonance with a Nitrogen-Vacancy Spin Sensor. *Science* **339**, 557–560 (2013).
11. Müller, C. *et al.* Nuclear magnetic resonance spectroscopy with single spin sensitivity. *Nat Commun* **5**, 4703 (2014).
12. Sushkov, A. O. *et al.* Magnetic Resonance Detection of Individual Proton Spins Using Quantum Reporters. *Phys. Rev. Lett.* **113**, 197601 (2014).
13. Sasaki, K., Watanabe, H., Sumiya, H., Itoh, K. M. & Abe, E. Detection and control of single proton spins in a thin layer of diamond grown by chemical vapor deposition. *Applied Physics Letters* **117**, 114002 (2020).
14. DeVience, S. J. *et al.* Nanoscale NMR spectroscopy and imaging of multiple nuclear species. *Nature Nanotech* **10**, 129–134 (2015).
15. Glenn, D. R. *et al.* High-resolution magnetic resonance spectroscopy using a solid-state spin sensor. *Nature* **555**, 351–354 (2018).
16. Bucher, D. B., Glenn, D. R., Park, H., Lukin, M. D. & Walsworth, R. L. Hyperpolarization-Enhanced NMR Spectroscopy with Femtomole Sensitivity Using Quantum Defects in Diamond. *Phys. Rev. X* **10**, 021053 (2020).
17. Bruckmaier, F. *et al.* Imaging local diffusion in microstructures using NV-based pulsed field gradient NMR. *Science Advances* **9**, eadh3484 (2023).
18. Neuling, N. R., Allert, R. D. & Bucher, D. B. Prospects of single-cell nuclear magnetic resonance spectroscopy with quantum sensors. *Current Opinion in Biotechnology* **83**, 102975 (2023).
19. Liu, K. S. *et al.* Surface NMR using quantum sensors in diamond. *Proceedings of the National Academy of Sciences* **119**, e2111607119 (2022).

20. Liu, K. S. et al. Using Metal–Organic Frameworks to Confine Liquid Samples for Nanoscale NV-NMR. *Nano Lett.* **22**, 9876–9882 (2022).
21. Allert, R. D., Briegel, K. D. & Bucher, D. B. Advances in nano- and microscale NMR spectroscopy using diamond quantum sensors. *Chem. Commun.* **58**, 8165–8181 (2022).
22. Rizzato, R., von Grafenstein, N. R. & Bucher, D. B. Quantum sensors in diamonds for magnetic resonance spectroscopy: Current applications and future prospects. *Applied Physics Letters* **123**, 260502 (2023).
23. Freire-Moschovitis, F. A. et al. The Role of Electrolytes in the Relaxation of Near-Surface Spin Defects in Diamond. *ACS Nano* **17**, 10474–10485 (2023).
24. Staudacher, T. et al. Probing molecular dynamics at the nanoscale via an individual paramagnetic centre. *Nat Commun* **6**, 8527 (2015).
25. London, P. et al. Detecting and Polarizing Nuclear Spins with Double Resonance on a Single Electron Spin. *Phys. Rev. Lett.* **111**, 067601 (2013).
26. Tetienne, J.-P. et al. Prospects for nuclear spin hyperpolarization of molecular samples using nitrogen-vacancy centers in diamond. *Phys. Rev. B* **103**, 014434 (2021).
27. Rizzato, R., Bruckmaier, F., Liu, K. S., Glaser, S. J. & Bucher, D. B. Polarization Transfer from Optically Pumped Ensembles of N-$V$ Centers to Multinuclear Spin Baths. *Phys. Rev. Applied* **17**, 024067 (2022).
28. Broadway, D. A. et al. Quantum probe hyperpolarisation of molecular nuclear spins. *Nat Commun* **9**, 1246 (2018).
29. Álvarez, G. A. et al. Local and bulk 13C hyperpolarization in nitrogen-vacancy-centred diamonds at variable fields and orientations. *Nat Commun* **6**, 8456 (2015).
30. Healey, A. J. et al. Polarization Transfer to External Nuclear Spins Using Ensembles of Nitrogen-Vacancy Centers. *Phys. Rev. Applied* **15**, 054052 (2021).
31. Eills, J. et al. Spin Hyperpolarization in Modern Magnetic Resonance. *Chem. Rev.* **123**, 1417–1551 (2023).
32. Ofori-Okai, B. K. et al. Spin properties of very shallow nitrogen vacancy defects in diamond. *Phys. Rev. B* **86**, 081406 (2012).
33. Mindarava, Y. et al. Efficient conversion of nitrogen to nitrogen-vacancy centers in diamond particles with high-temperature electron irradiation. *Carbon* **170**, 182–190 (2020).
34. Vaidya, S., Gao, X., Dikshit, S., Aharonovich, I. & Li, T. Quantum sensing and imaging with spin defects in hexagonal boron nitride. *Advances in Physics: X* **8**, 2206049 (2023).
35. Montblanch, A. R.-P., Barbone, M., Aharonovich, I., Atatüre, M. & Ferrari, A. C. Layered materials as a platform for quantum technologies. *Nat. Nanotechnol.* **18**, 555–571 (2023).
36. Caldwell, J. D. et al. Photonics with hexagonal boron nitride. *Nat Rev Mater* **4**, 552–567 (2019).
37. Cassabois, G., Valvin, P. & Gil, B. Hexagonal boron nitride is an indirect bandgap semiconductor. *Nature photonics* **10**, 262–266 (2016).
38. Aharonovich, I., Tetienne, J.-P. & Toth, M. Quantum Emitters in Hexagonal Boron Nitride. *Nano Lett.* **22**, 9227–9235 (2022).
39. Cholsuk, C., Zand, A., Çakan, A. & Vogl, T. The hBN Defects Database: A Theoretical Compilation of Color Centers in Hexagonal Boron Nitride. *J. Phys. Chem. C* **128**, 12716–12725 (2024).
40. Gottscholl, A. et al. Initialization and read-out of intrinsic spin defects in a van der Waals crystal at room temperature. *Nature Materials* **19**, 540–545 (2020).
41. Stern, H. L. et al. Room-temperature optically detected magnetic resonance of single defects in hexagonal boron nitride. *Nature communications* **13**, 1–9 (2022).
42. Stern, H. L. et al. A quantum coherent spin in hexagonal boron nitride at ambient conditions. *Nature Materials* (2024) doi:10.1038/s41563-024-01887-z.
43. Mendelson, N. et al. Identifying carbon as the source of visible single-photon emission from hexagonal boron nitride. *Nature Materials* **20**, 321–328 (2021).
44. Liu, W. et al. Spin-active defects in hexagonal boron nitride. *Materials for Quantum Technology* **2**, 032002 (2022).
45. Durand, A. et al. Optically Active Spin Defects in Few-Layer Thick Hexagonal Boron Nitride. *Phys. Rev. Lett.* **131**, 116902 (2023).
46. Baber, S. et al. Excited State Spectroscopy of Boron Vacancy Defects in Hexagonal Boron Nitride Using Time-Resolved Optically Detected Magnetic Resonance. *Nano Lett.* **22**, 461–467 (2022).

47. Gottscholl, A. *et al.* Room temperature coherent control of spin defects in hexagonal boron nitride. *Science Advances* **7**, eabf3630.
48. Gottscholl, A. *et al.* Spin defects in hBN as promising temperature, pressure and magnetic field quantum sensors. *Nature Communications* **12**, 4480 (2021).
49. Haykal, A. *et al. Decoherence of VB- Spin Defects in Monoisotopic Hexagonal Boron Nitride*. https://www.researchsquare.com/article/rs-1185884/v1 (2022) doi:10.21203/rs.3.rs-1185884/v1.
50. Gao, X. *et al.* Nuclear spin polarization and control in hexagonal boron nitride. *Nature Materials* (2022) doi:10.1038/s41563-022-01329-8.
51. Gong, R. *et al.* Coherent dynamics of strongly interacting electronic spin defects in hexagonal boron nitride. *Nature Communications* **14**, 3299 (2023).
52. Huang, M. *et al.* Wide field imaging of van der Waals ferromagnet Fe3GeTe2 by spin defects in hexagonal boron nitride. *Nature Communications* **13**, 5369 (2022).
53. Rizzato, R. *et al.* Extending the coherence of spin defects in hBN enables advanced qubit control and quantum sensing. *Nature Communications* **14**, 5089 (2023).
54. Healey, A. J. *et al.* Quantum microscopy with van der Waals heterostructures. *Nat. Phys.* **19**, 87–91 (2023).
55. Cohen, D., Nigmatullin, R., Eldar, M. & Retzker, A. Confined Nano-NMR Spectroscopy Using NV Centers. *Advanced Quantum Technologies* **3**, 2000019 (2020).
56. Blankenship, B. W. *et al.* Complex Three-Dimensional Microscale Structures for Quantum Sensing Applications. *Nano Lett.* **23**, 9272–9279 (2023).
57. Gierse, M. *et al.* Scalable and Tunable Diamond Nanostructuring Process for Nanoscale NMR Applications. *ACS Omega* **7**, 31544–31550 (2022).
58. Kehayias, P. *et al.* Solution nuclear magnetic resonance spectroscopy on a nanostructured diamond chip. *Nat Commun* **8**, 188 (2017).
59. Holzgrafe, J. *et al.* Nanoscale NMR Spectroscopy Using Nanodiamond Quantum Sensors. *Phys. Rev. Applied* **13**, 044004 (2020).
60. Gao, X. *et al.* Nanotube spin defects for omnidirectional magnetic field sensing. *Nature Communications* **15**, 7697 (2024).
61. Allard, C. *et al.* Confinement of Dyes inside Boron Nitride Nanotubes: Photostable and Shifted Fluorescence down to the Near Infrared. *Advanced Materials* **32**, 2001429 (2020).
62. Cohen, M. L. & Zettl, A. The physics of boron nitride nanotubes. *Physics Today* **63**, 34–38 (2010).
63. Jordan, J. W. *et al.* Host–Guest Chemistry in Boron Nitride Nanotubes: Interactions with Polyoxometalates and Mechanism of Encapsulation. *J. Am. Chem. Soc.* **145**, 1206–1215 (2023).
64. Zhi, C., Bando, Y., Tang, C. & Golberg, D. Immobilization of Proteins on Boron Nitride Nanotubes. *J. Am. Chem. Soc.* **127**, 17144–17145 (2005).
65. Liu, K. S. *et al.* Using Metal–Organic Frameworks to Confine Liquid Samples for Nanoscale NV-NMR. *Nano Lett.* (2022) doi:10.1021/acs.nanolett.2c03069.
66. Schwartz, I. *et al.* Robust optical polarization of nuclear spin baths using Hamiltonian engineering of nitrogen-vacancy center quantum dynamics. *Sci. Adv.* **4**, eaat8978 (2018).
67. McGuinness, L. P. *et al.* Quantum measurement and orientation tracking of fluorescent nanodiamonds inside living cells. *Nature Nanotech* **6**, 358–363 (2011).
68. Doherty, M. W. *et al.* The nitrogen-vacancy colour centre in diamond. *Physics Reports* **528**, 1–45 (2013).
69. Sarkar, A. *et al.* High-precision chemical quantum sensing in flowing monodisperse microdroplets. *Science Advances* **10**, eadp4033.
70. Scholten, S. C. *et al.* Multi-species optically addressable spin defects in a van der Waals material. *Nature Communications* **15**, 6727 (2024).
71. Guo, N.-J. *et al.* Coherent control of an ultrabright single spin in hexagonal boron nitride at room temperature. *Nature Communications* **14**, 2893 (2023).
72. Singh, P. *et al.* Violet to near-infrared optical addressing of spin pairs in hexagonal boron nitride. Preprint at http://arxiv.org/abs/2409.20186 (2024).
73. Robertson, I. O. *et al.* A universal mechanism for optically addressable solid-state spin pairs. *arXiv preprint arXiv:2407.13148* (2024).

74. Robertson, I. O. *et al.* Detection of Paramagnetic Spins with an Ultrathin van der Waals Quantum Sensor. *ACS Nano* **17**, 13408–13417 (2023).
75. Gao, X. *et al.* Quantum Sensing of Paramagnetic Spins in Liquids with Spin Qubits in Hexagonal Boron Nitride. *ACS Photonics* **10**, 2894–2900 (2023).
76. Carr, H. Y. & Purcell, E. M. Effects of Diffusion on Free Precession in Nuclear Magnetic Resonance Experiments. *Phys. Rev.* **94**, 630–638 (1954).
77. Meiboom, S. & Gill, D. Modified spin-echo method for measuring nuclear relaxation times. *Review of scientific instruments* **29**, 688–691 (1958).
78. Bar-Gill, N. *et al.* Suppression of spin-bath dynamics for improved coherence of multi-spin-qubit systems. *Nat Commun* **3**, 858 (2012).
79. Sousa, R. de. Electron Spin as a Spectrometer ofáNuclear-SpináNoiseáand Other Fluctuations. in *Electron spin resonance and related phenomena in low-dimensional structures* 183–220 (Springer, 2009).
80. Ryan, C. A., Hodges, J. S. & Cory, D. G. Robust decoupling techniques to extend quantum coherence in diamond. *Physical Review Letters* **105**, 200402 (2010).
81. Schweiger, A. & Jeschke, G. *Principles of Pulse Electron Paramagnetic Resonance*. (Oxford University Press on Demand, 2001).
82. Hermann, J. C. *et al.* Extending radiowave frequency detection range with dressed states of solid-state spin ensembles. *npj Quantum Inf* **10**, 1–7 (2024).
83. Henstra, A., Dirksen, P., Schmidt, J. & Wenckebach, W. T. Nuclear spin orientation via electron spin locking (NOVEL). *Journal of Magnetic Resonance (1969)* **77**, 389–393 (1988).
84. Steinert, S. *et al.* Magnetic spin imaging under ambient conditions with sub-cellular resolution. *Nature Communications* **4**, 1607 (2013).
85. Simpson, D. A. *et al.* Electron paramagnetic resonance microscopy using spins in diamond under ambient conditions. *Nature Communications* **8**, 458 (2017).
86. Ziem, F. C., Götz, N. S., Zappe, A., Steinert, S. & Wrachtrup, J. Highly Sensitive Detection of Physiological Spins in a Microfluidic Device. *Nano Lett.* **13**, 4093–4098 (2013).
87. Nie, L. *et al.* Quantum monitoring of cellular metabolic activities in single mitochondria. *Science Advances* **7**, eabf0573.
88. Nie, L. *et al.* Quantum Sensing of Free Radicals in Primary Human Dendritic Cells. *Nano Lett.* **22**, 1818–1825 (2022).
89. Han, N. *et al.* Rational design of boron nitride with different dimensionalities for sustainable applications. *Renewable and Sustainable Energy Reviews* **170**, 112910 (2022).
90. Allert, R. D. *et al.* Microfluidic quantum sensing platform for lab-on-a-chip applications. *Lab Chip* **22**, 4831–4840 (2022).
91. Ajoy, A. *et al.* Orientation-independent room temperature optical 13C hyperpolarization in powdered diamond. *Science Advances* **4**, eaar5492.
92. Fernández-Acebal, P. *et al.* Toward Hyperpolarization of Oil Molecules via Single Nitrogen Vacancy Centers in Diamond. *Nano Lett.* **18**, 1882–1887 (2018).
93. Bucher, D. B. *et al.* Quantum diamond spectrometer for nanoscale NMR and ESR spectroscopy. *Nat Protoc* **14**, 2707–2747 (2019).